\documentclass[referee]{aa}
\usepackage{graphicx}
\usepackage{longtable}
\usepackage{txfonts}
\usepackage{lscape}
\usepackage{threeparttable}

\begin{document}

\title{Long-Time Evolution of Gas-Free Disc Galaxies in Binary Systems}

\author{R. Chan \inst{1} \and S. Junqueira \inst{2}}
\institute{Coordena\c c\~ao de Astronomia e Astrof\'{\i}sica,
Observat\'orio Nacional, Rua General Jos\'e Cristino 77, S\~ao
Crist\'ov\~ao, CEP 20921--400, Rio de Janeiro, RJ, Brazil. 
\email{chan@on.br}
\and
Divis\~ao Servi\c co da Hora,
Observat\'orio Nacional, Rua General Jos\'e Cristino 77, S\~ao
Crist\'ov\~ao, CEP 20921--400, Rio de Janeiro, RJ, Brazil.
\email{selma@on.br}
}

\authorrunning{R. Chan \inst{1} \and S. Junqueira \inst{2}}
\titlerunning{Long-Time Evolution of Gas-Free Disc Galaxies}

\date{Received ; Accepted}

\maketitle

\begin{abstract}
~We present the results of several detailed numerical N-body simulations of the 
dynamical interactions of two equal mass disc galaxies. Both galaxies are 
embedded in spherical halos of dark matter and contain central bulges. Our 
analysis of the dynamical evolution of the binary system focuses on the 
morphological evolution of the stellar distribution of the discs. 
The satellite galaxy has coplanar or polar disc orientation in relation 
to the disc of the primary galaxy and their initial orbits are prograde
eccentric ($e=0.1$, $e=0.4$ or $e=0.7$).  Both galaxies have mass and size 
comparable to the Milky Way. 
We show that the 
merger of the two disc galaxies, depending on the relative orientation of the 
discs, can yield either a disc or lenticular remnant, instead of an elliptical 
one. These are the first simulations in the literature to show the formation of 
S0-like galaxies from protracted binary galaxy interactions. Additionally, we 
demonstrate that the time to merger increases linearly with the initial 
apocentric distance between the galaxies, and decreases with the initial 
orbital eccentricity. We also show that the tidal forces of the discs excite 
transient $m=1$ and $m=2$ wave modes, i.e., lopsidedness, spiral arms, and bars.
However, after the merging of the discs, such instabilities fade completely, 
and the remnant is thicker and bigger than the original discs.
The maximum relative amplitude of these waves is at most about 15 times greater 
compared to the control case. 
The $m=2$ wave mode is generated mainly by tidal interaction in the outer 
region of 
the discs. The $m=1$ wave mode depends mostly of an interaction of the
inner part of the discs, producing an off-centering effect of the wave mode
center relative to the center of 
mass of the disc. These characteristics produce a time lag among the maximum
formation
of these two wave modes. 
Finally, the disc settles down quickly, after the 
merger, in less than one outer disc rotation period.
\end{abstract}

\keywords{
Simulation -- spiral 
-- lopsidedness -- binary galaxies -- merger -- lenticular galaxies}

\section{Introduction}

Seventy percent of galaxies in the nearby universe are
characterized by a disc with prominent spiral arms, but our
understanding of the origin of these patterns is incomplete, even
after decades of theoretical study \cite{Sellwood2011,Sellwood2013}. 
Several ideas have been proposed to explain the formation of spiral arms.

The latest simulations show that gravitational
instabilities in the stars lead to flocculent and multi-armed spirals
which persist for many Gyrs \cite{Oh2008,Fujii2011}.
However, the mechanism which produces and maintains two-armed
grand design galaxies is still ambiguous. 

Grand design galaxies, which exhibit symmetric two-armed
spiral structures, represent a significant fraction of spiral galaxies.
{ The production of such a spiral} galaxy faces two major obstacles: first,
inducing the m = 2 spiral structure, and secondly maintaining it.

It is known that the spiral arms of disc galaxies can be
excited by tidal interactions with nearby companion galaxies
\cite{Oh2008,Dobbs2010,Struck2011}.

Oh and collaborators \cite{Oh2008} have investigated the physical properties of 
tidal structures in a disc galaxy created by gravitational interactions with
a companion using numerical N-body simulations. They have considered a  
galaxy model consisting of a rigid halo/bulge and an infinitesimally thin 
stellar disc with Toomre parameter $Q\approx 2$. The perturbing companion was 
treated as a pointmass moving on a prograde parabolic orbit, with varying mass 
and pericenter distance. They have shown that tidal interactions produce 
well-defined spiral arms and extended tidal features, such as bridge and tail, 
which are all transients.

Dobbs's et al. (2010) modeled the disc galaxy M51 and its 
interaction with a companion
point-mass NGC 5195, focusing primarily on the dynamics of the gas,
and secondly the stellar disc. 
The halo was represented by a rigid potential.
The tidal interaction has produced spiral
arms in the stars and in the gas. The resulting spiral structure has shown
excellent agreement with that of M51.

{ The work of Lotz et al. (2010) analyzed} the effect of a gas 
fraction on the morphologies of a series of simulated disc galaxy mergers.
Each galaxy was initially modeled as a disc of
stars and gas, a stellar bulge and a dark matter halo, with different number
of particles and masses for each component.
All the simulated mergers had the same orbital parameters. Each pair of 
galaxies has started on a sub-parabolic orbit with eccentricity 0.95 and an 
initial 
pericentric radius of 13.6 kpc. The galaxies have had a roughly 
prograde-prograde 
orientation relative to the orbital plane, with the primary galaxy tilted 
30$^\circ$ from a pure prograde orientation.
Their simulations have predicted that galaxy mergers would exhibit high
asymmetries for longer periods of time if they have had high gas fractions.

Struck and collaborators \cite{Struck2011} have discovered long-lived waves in 
numerical simulations of fast (marginally bound or unbound) flyby galaxy 
collisions. The main galaxy has had a rigid halo potential, gas and
the companion was modeled as a point mass.  They have found that none of the 
simulations has resulted in bar formation.
They have also shown that while these waves propagate through the disc, they 
are maintained by the coherent oscillations initiated by the impulsive 
disturbance.

Snaith at al. (2012) have studied the properties and evolution 
of a simulated polar disc galaxy. This galaxy was composed of two orthogonal 
discs, one of which contains old stars (old stellar disc) and the
other both younger stars and cold gas (polar disc).
They have confirmed that the polar disc galaxy is the result of the last major 
merger, where the angular moment of the interaction is orthogonal to
the angle of the infalling gas.

{ In one of our previous works in kinematic and morphology of spiral 
galaxies we have shown} a deep interaction between the dynamical and
morphological properties of this kind of galaxy \cite{Chan2003}.
With continual satellite forcing, the final state was in 
the form of a slowly evolving wave pattern, as shown by the existence of
pattern speeds for stable $m=1$ and $m=2$ wave modes.
The pattern speeds obtained from the density and the
three positive velocity component distributions are the
same. This was also true for the negative velocity components.

Kinematic studies of spiral galaxies have revealed a remarkable 
variety of
interesting behavior: some galaxies have large scale asymmetries in their
rotation curves as signature of kinematic lopsidedness 
\cite{Junqueira1996},
while in others the inner regions counter-rotate with the respect to the rest
of the galaxy \cite{Garcia-Burillo2000}.  
Most of the spiral galaxies have asymmetric HI profiles and asymmetric 
rotation curves \cite{Haynes2000,Andersen2013}. 
Such intriguing kinematics could plausibly result if these galaxies
are the end-products of minor mergers \cite{Haynes2000}.  Minor mergers 
and weak tidal interactions between galaxies occur with much higher frequency
than major ones.  By weak interactions between
galaxies we mean those that do not destroy the disc of the primary galaxy.
However, weak interactions may cause disc heating and 
satellite remnants may build up the stellar halo.  

Galaxies' interactions are likely to play a key role in determining the
morphology and the dynamical properties of disc galaxies.
Careful examination shows that most disc galaxies are not truly symmetric
but exhibit a variety of morphological peculiarities of which spiral arms 
and bars are the most pronounced. 
Disc galaxies currently show significant
spiral-generating tidal perturbations by one or more small-mass
companions, and nearly all have had tidal
interactions at sometime in the past.  

After decades of efforts, we now know that these
features may be driven by environmental disturbances acting directly on the
disc, in addition to self-excitation of a local disturbance (e.g., by swing
amplification){ \cite{Toomre1981}}. However, the disc is embedded within 
a halo and, therefore,
the luminous disc is not dynamically independent \cite{Combes2008}. The dark 
matter halo is
disturbed by dwarf companions, neighboring galaxies in groups and clusters and
the tidal force from the overall cluster.
If the halo can respond globally to such
disturbances, it can affect the disc structure.
Thus, because most
spirals have dwarf companions, interactions with these companions are
present and the inward propagation of external
perturbations by the halo could be a dominant source of disc structure for 
all galaxies \cite{Vesperini2000}.

In order to study the dynamical evolution of two 
disc galaxies and their morphological evolution, this paper explores the 
picture 
as follows.  First, we assume a disc galaxy with the characteristics of the
Milk Way (disc, bulge and halo).  { Second, we let a secondary galaxy 
orbit on prograde coplanar or polar
disc (orientation in relation to the primary disc galaxy). Although the gas is 
important in modeling a
realistic disc} galaxy, in this work we focus our attention only to the
morphological stellar properties.

As an example of work in evolution of stellar disc galaxies without gas, we can 
mention the recent published work by Baba, Saitoh \& Wada \cite{Baba2013}. Using
N-body simulations, they analyzed the physical mechanisms of non steady stellar
spiral arms in disk galaxies, i.e., without the gaseous component. They have
studied the growing and damping phases of the spiral arms and they have
confirmed that the spiral arms are formed due to a swing amplification
mechanism that reinforces density enhancement.  The main motivation was that
all the previous time-dependent simulation works have not been able to prove
the existence of stationary density waves in a disk galaxy without external
perturbations and a bar structure.

{ Thus, the main goal of the present work} is, utilizing detailed numerical 
N-body 
simulations, to study the dynamical interactions of the two discs of the 
galaxies.
In particular, we have investigated 
whether interactions can induce a persistent and stable 
$m=1$ or $m=2$ patterns in disc galaxies.

The paper is  organized as follows: in Section 2 we describe the numerical 
method used in the simulations.
In Section 3 we present the initial conditions.  In Section 4 we show
power spectra of the instabilities.  
Finally, in Section 5 we discuss and summarize the results.

\section {The Numerical Method}

The full N-body code utilized in the simulations was GADGET \cite{Springel2001}.
The code was parallelized and the communication is done
by means of the Message Passing Interface (MPI).
GADGET evolves self-gravitating collisionless fluids with the traditional 
N-body approach, and a collisional gas by smoothed particle hydrodynamics. 
But in our case we use only the particle integration, which uses
a tree algorithm to compute gravitational forces.
The parallel version has been designed to run on massively parallel 
supercomputers with distributed memory. 

Fortin, Athanassoula \& Lambert (2011) published a comparison with different
codes for galactic N-body simulations, in particular, the GADGET code and the
Dehnen's algorithm.  They have shown that the serial implementation of the
Dehnen's algorithm is slower, in terms of execution time, in comparison with
the parallelized implementation of the GADGET code, with 8 to 128 processors.

The GADGET code does not exactly conserve energy or momentum, but the energy is
conserved to better than 0.3\% over an entire evolution, and the center of the
mass moves a distance of at most $82 \epsilon$ (the softening parameter)
from the initial system center mass, with a time step size
$\Delta t = 1.000\times 10^{-3}$ and $\epsilon=8.000\times10^{-4}$.
Too large $\epsilon$ reduces the noise but it increases the error in the
calculation of the force due to the failure to resolve the interactions of
particles with scale lengths less than it.
On the other hand, too small value for $\epsilon$ produces a noisy estimation
of the force due to the finite number of particles.
The value of the optimal softening depends both
on the mass distribution and on the number of particles used to represent it.
For a higher number of particles the optimal softening is smaller.
More concentrated mass distributions necessitate smaller softening.
Several works have already analyzed this problem carefully \cite{Merrit1996}
\cite{Athanassoula2000}\cite{Rodionov2005}.  For example, Athanassoula et al. 
(2000) have shown that, for a Dehnen sphere, $\epsilon=4.000\times10^{-3}$.
Thus, the softening scale is motivated by a tradeoff between accuracy and
computational speed.
We have run several simulations of the isolated disc
galaxy in order to find the optimal $\epsilon$ and maximum integration time 
step. The criteria were to find the maximum softening parameter and integration
time step such that it could minimize the heating of the disk, without changing
too much the physical disc parameters, mainly $Z_d$.  When the $\epsilon$ was
too big then the disc heated up, increasing the width of it.
Thus, the chosen softening parameter is smaller than the disc scale height
$Z_d$, in order to the disk of the galaxies being followed accurately.

For non zero tolerance parameter $\theta$, the treecode CPU time scales as 
$O(N \log N)$, but in the
limit $\theta \rightarrow 0$ this method scales as $O(N^2)$, approaching to the
direct code.  At the very beginning of this work we have used the GADGET-1
\cite{Springel2001}.
Thus, we have assumed for the tolerance parameter $\theta$ the value 0.577,
in order to avoid pathological situations when using the treecode as described
by Salmon \& Warren \cite{Salmon1994}.  However, this value of $\theta$ is CPU
time-consuming but the error in the force calculation is minimized, although
many authors studying disc galaxies have used $\theta > 0.7$ \cite{Oh2008}
\cite{Minchev2012}.

\section{The Initial Conditions of the Simulations}

We have used in the simulations the nearly self-consistent disc-bulge-halo 
galaxy model proposed by Kuijken \& Dubinski \cite{Kuijken1995}.  We assumed a 
model that has 
mass distributions and rotation curves closely resembling of the Milk Way,
i.e., the model MW-A of the Kuijken \& Dubinski's work \cite{Kuijken1995}.
This galaxy disc model has a disc-bulge mass ratio of 2:1
and halo-disc mass ratio of 5:1 (see Table 
\ref{table1}). The disc is warm with a Toomre parameter $Q=1.7$ at the disc
half-mass radius.

Our simulations are based on a small number of particles in comparison with
several works in disc galaxies.  However, since none of the works in the
literature has tackled before this kind of problem proposed here, we have
decided to run modest simulations using the computer clusters that were
available, in order to have, at least, 
{ an initial dynamical study of these kinds of binary galaxies}.

\begin{table*}
\begin{minipage}{160 mm}
\caption{Disc Galaxy Model Properties}
\label{table1}
\begin{tabular}{@{}cccccccccccc}
Galaxy & $M_d$ & $N_d$ & $R_d$ & $Z_d$ & $R_t$ & $M_b$ & $N_b$ & $M_h$ & $N_h$ & $m$ & $\epsilon$ \\
$G_1$ & 0.871 & 40,000 & 1.000 & 0.100 & 5.000 & 0.425 & 19,538 & 4.916 & 225,880 & $2.176\times10^{-5}$ & $8.000\times10^{-4}$ \\
\end{tabular}

\medskip
$M_d$ is the disc mass in units of mass, $N_d$ the number of particles of the disc, 
$R_d$ the disc scale radius in units of length, $Z_d$ the disc scale height in units of length,
$R_t$ the disc truncation radius in units of length,
$M_b$ the bulge mass in units of mass, $N_b$ the number of particles in the bulge, 
$M_h$ the halo mass in units of mass, $N_h$ the number of halo particles,
$m$ the mass of each particle in units of mass and
$\epsilon$ the softening of each particle in units of length.
\end{minipage}
\end{table*}

The disc follows approximately an exponential-sech law described by
\begin{equation}
\rho_d(R,Z) =
\rho_o \exp \left( -\frac{R}{R_d} \right) {\rm sech}^2 
\left( \frac{Z}{Z_d} \right),
\label{dens}
\end{equation}
where $\rho_o$ is the central density that is related to the total mass of
the disc. This approximation has been used because the full potential 
equation obtained
by Kuijken \& Dubinski \cite{Kuijken1995} is analytically more complicated.

\section{The Results of the Simulations}

\begin{figure}
\centering
\includegraphics[width=8cm]{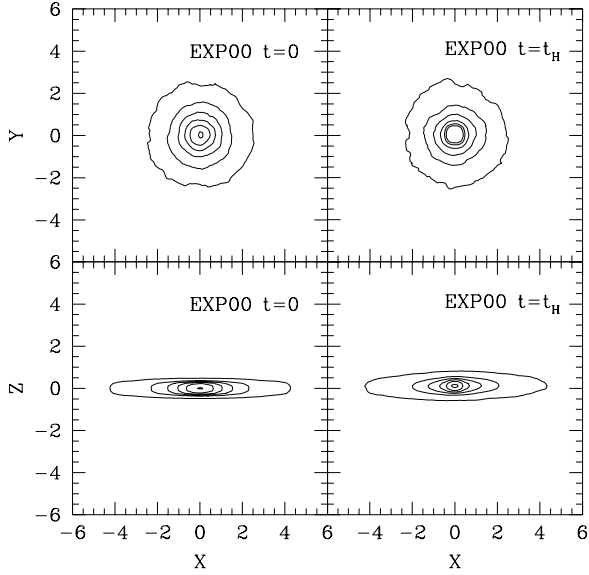}
\caption
{The contour plot of the primary galaxy at the beginning of the 
simulation ($t=0$) and at the Hubble time of the simulation ($t=t_H$).
The smoothing was done, averaging the 25 first and second neighbors of
each pixel. Hereinafter, the density levels in the planes XY and XZ at $t=0$ 
will be used in all the contour plots, in the planes XY and XZ, respectively.}
\label{snapshot_000}
\end{figure}

In the Figure \ref{snapshot_000} we show the contour plot of the primary 
galaxy at the beginning of the
simulation ($t=0$) and at the Hubble time of the simulation ($t=t_H$).
We note that the central density in the plane XY has increased slightly
after one
Hubble time of simulation, since the contour levels are the same for the
two instants of time. In the XZ plane the scale height apparently
has increased due to the 2-body relaxation heating, however we can see that
this quantity has changed very little (see Figure \ref{Zd}).

\begin{figure}
\centering
\includegraphics[width=8cm]{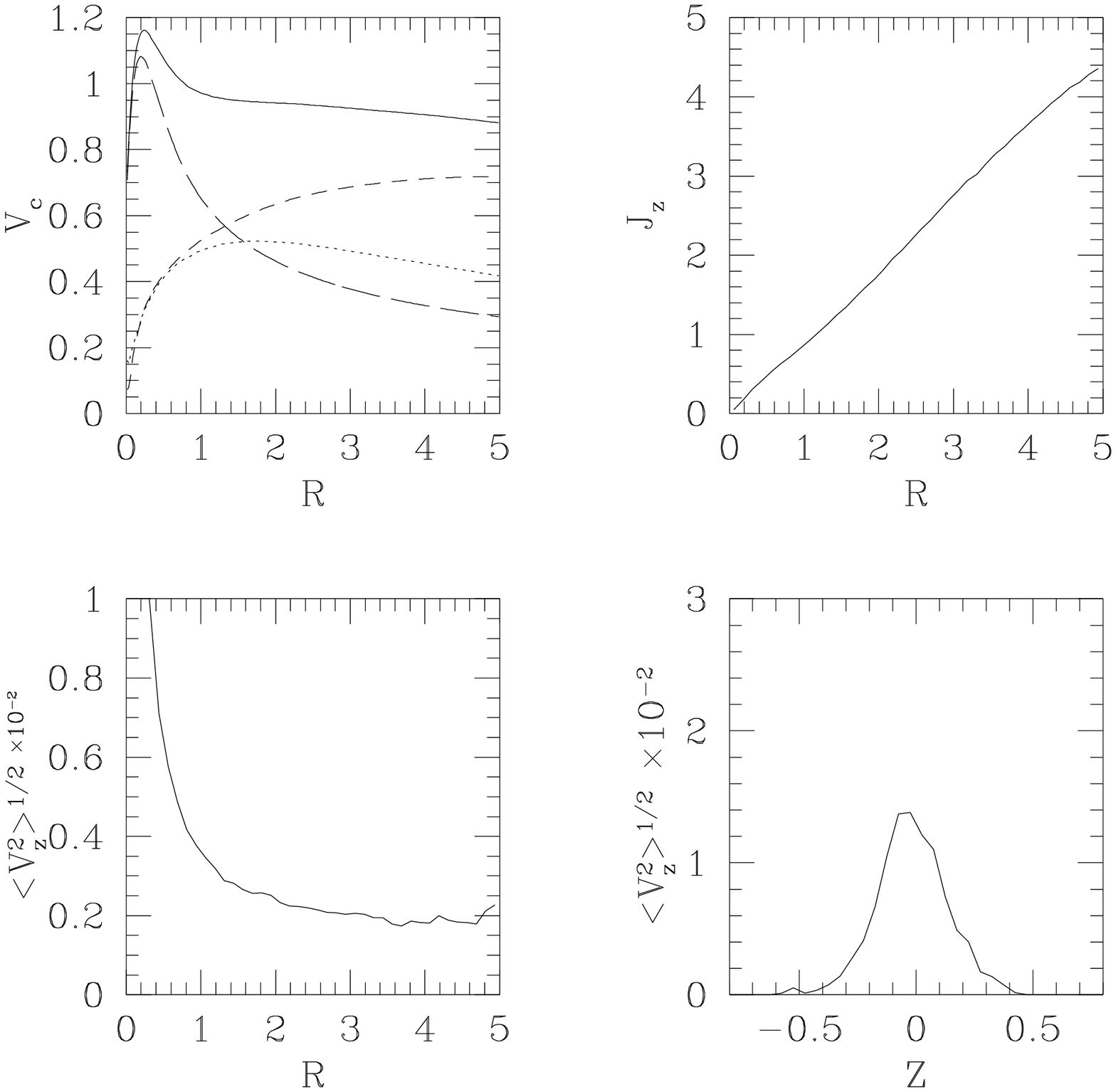}
\caption
{The rotation curve at the time $t=0$ of the disc $V_c$,
the main component of the angular momentum per unit of mass $J_z$ and the
velocity dispersion in the $Z$ direction $<V^2_z>^{1/2}$.
The coordinate $R$ is the  radius in cylindrical coordinates. The dotted line
denotes the disc, the long-dashed line denotes the bulge, the short-dashed
line denotes the halo and the solid line denotes the total rotation curve.}
\label{rotcurve_000}
\end{figure}

\begin{figure}
\centering
\includegraphics[width=8cm]{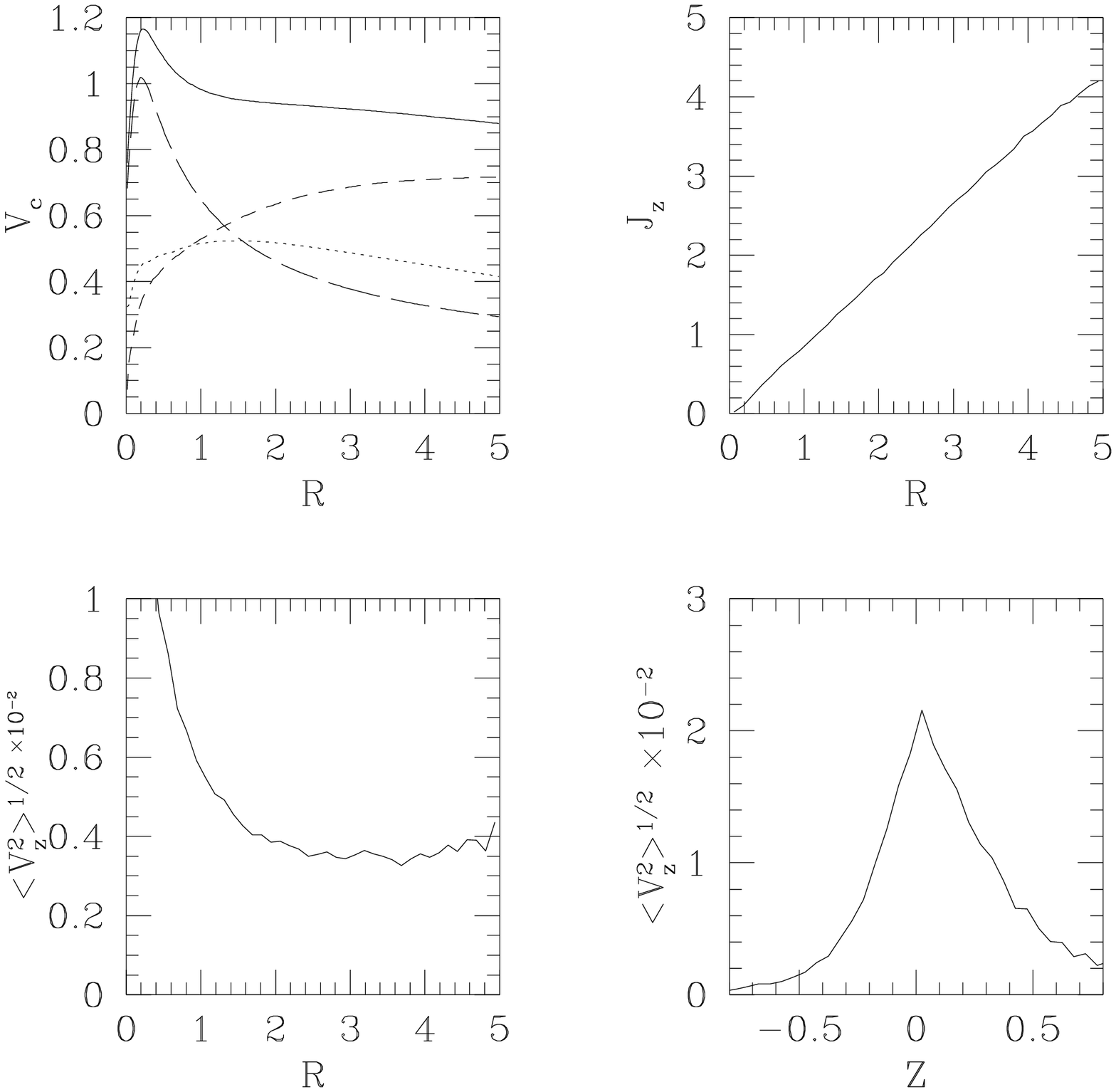}
\caption
{The rotation curve at the time $t=t_H$ of the disc $V_c$,
the main component of the angular momentum per unit of mass $J_z$ and the
velocity dispersion in the $Z$ direction $<V^2_z>^{1/2}$.
The coordinate $R$ is the  radius in cylindrical coordinates. The dotted line
denotes the disc, the long-dashed line denotes the bulge, the short-dashed
line denotes the halo and the solid line denotes the total rotation curve.}
\label{rotcurve_100}
\end{figure}

\begin{figure}
\centering
\includegraphics[width=8cm]{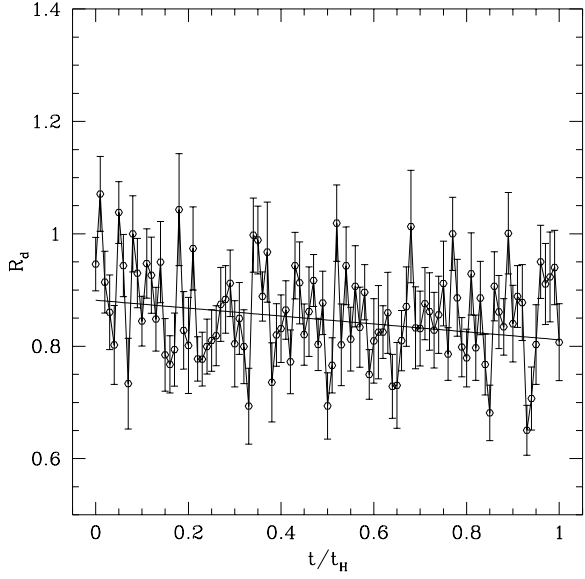}
\caption
{The time evolution of the scale radius ($R_d$).
The projected particle number density on the XY plane was fitted using the
approximation given by the Equation (\ref{dens}).
The coordinate $R$ is the  radius in cylindrical coordinates.
The fitting parameters are: $R_d=(-0.7042\times 10^{-1}\pm 0.2840\times 
10^{-1})[t/t_H]$ + $(0.8819\pm 0.1620\times 10^{-1})$.} 
\label{Rd}
\end{figure}

\begin{figure}
\centering
\includegraphics[width=8cm]{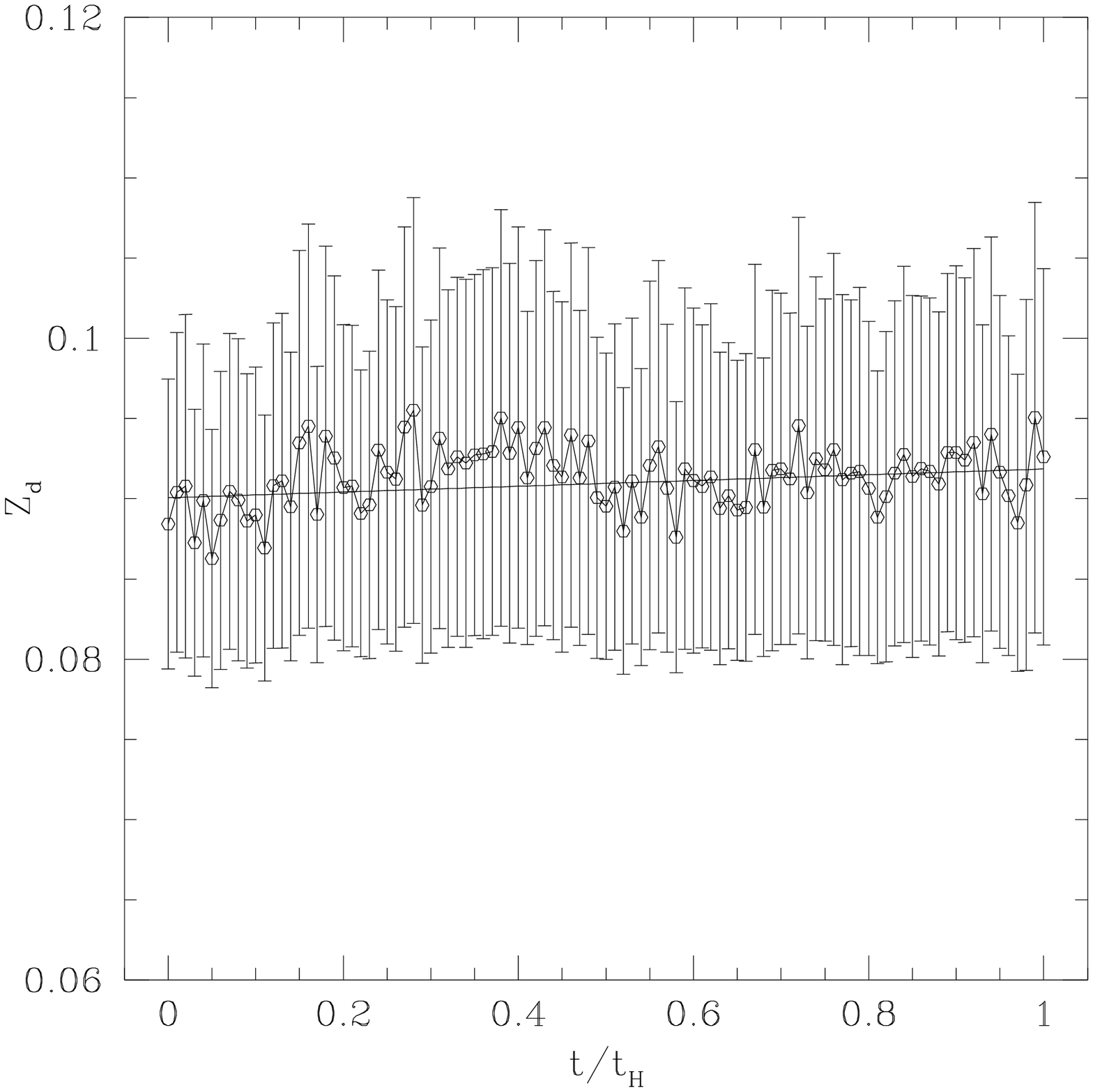}
\caption
{The time evolution of the scale height ($Z_d$).
The projected particle number density on the XZ plane was fitted using the
approximation given by the Equation (\ref{dens}).
The fitting parameters are: $Z_d=(0.1791\times 10^{-2}\pm 0.6320\times 
10^{-3})[t/t_H]$ + $(0.9006\times 10^{-1}\pm 0.3563\times 10^{-3})$.}
\label{Zd}
\end{figure}

Comparing the Figures \ref{rotcurve_000} and \ref{rotcurve_100} we note
from the quantity $<V_z^2>^{1/2}$ that
the self-heating of the initial disc and the particle halo add another
significant source of heating in the disc.  The gravitational 
softening can also cause the disc to puff up, this is the reason we have chosen
a such small softening parameter, 125 times smaller than the scalar disc 
height.  We can also observe that the total rotation curves $V_c$ and
the angular momentum in the Z direction have not changed, after one Hubble 
time of simulation. 

In the Figures \ref{Rd} and \ref{Zd} we present the time evolution of the
scale radius ($R_d$) and the scale height ($Z_d$).  We notice that,
as expected, due to
the heating of the disc the first quantity diminishes with the time
while the second increases with the time. The linear fitting parameters
of these two quantities are presented in the captions of these figures.
Since the scale height has increased less than 0.2\%, we have assumed,
hereinafter, that this
scale has not changed when we analyzed the data of the simulations.

The units used in the simulations are: $G=1$, [length] = $4.500$ kpc, [mass] =
$5.100\times 10^{10} M_\odot$, [time] = $1.993\times 10^{7}$ years 
($H_0 = 100$ km/s/Mpc) and [velocity] = 220.730 km/s.
Hereinafter, all the physical quantities will be referred in these units.
The particle softening radius was assumed to be 0.0008 or 1/125 of the disc 
scale height.  The critical opening angle was set to $\theta=0.577$ and the
forces between the cells and particles used the quadrupole correction.
The maximum integration step time was assumed to be 0.001 in units of 
simulation time.
The Hubble time $t_H$ corresponds to 490 time units.

We have run several simulations, with no satellite in order to
check the initial instabilities of the galaxy model.  The initial galaxy 
simulations
were run in a SUN FIRE 6800 cluster, with 16 CPU processors. Each simulation 
has taken about 50 days of CPU time.  For the simulations with the primary
and satellite galaxies we have used several clusters with a variety of CPU
processors: SUN BLADE X6250, SUN FIRE X2200, SGI ALTIX ICE 8200, 
SGI ALTIX 450/1350, SGI ALTIX-XE 340, IBM P750, INTEL PENTIUM QUAD CORE and 
INTEL PENTIUM DUAL CORE.  The number of CPU processors varied from the minimum 
of 8 to the maximum of 128. Each simulation has taken 90 days of CPU time in
average. 

A great number of the works in the literature 
\cite{Oh2008,Dobbs2010,Lotz2010,Struck2011,
Bois2011} simulating interaction between two disc galaxies have used initial
conditions as parabolic or hyperbolic orbits. However, none has studied the
bounded eccentric orbits in an interval of a Hubble time because is very
expensive in terms of CPU time machine. This cost is mainly due to the fact that
in this kind of simulation we must integrate adequately the internal dynamic
of the disc galaxy during a huge cosmological interval of time. 
We have decided not to use cosmologically consistent initial conditions from
publicly available simulations as done, e.g., by Ruszkowski \& Springel
\cite{Ruszkowski2009} for two reasons. First, we would like to know what
happened with the binary galaxies in bound orbits and, secondly, it would cost
much more (in number of experiments and in CPU time) if we have begun our
simulations from cosmological unbound galaxies to arrive to eccentric bound
galaxies.

All the initial conditions of the numerical experiments are presented in
Table \ref{table2}. The orbits of the initial galaxies
are eccentric ($e=0.1$, $0.4$ or $0.7$) and the orientations of the discs are
coplanar ($\Theta=0$) or polar 
($\Theta=90$) to each other.  
The simulations begin with the two galaxies at the apocentric positions.

\begin{table*}
\begin{minipage}{80 mm}
\caption{Primary and Secondary Galaxy Initial Conditions}
\label{table2}
\begin{tabular}{@{}cccccc}
EXP & $\Theta$ & $R_p$ & $e$ & $R_a$ & $V_a$ \\
00 &     &    &     &        &        \\
01 &   0 & 30 & 0.1 &  36.67 & 0.5521 \\
02 &   0 & 30 & 0.4 &  70.00 & 0.3263 \\
03 &   0 & 30 & 0.7 & 170.00 & 0.1480 \\
04 &   0 & 40 & 0.1 &  48.89 & 0.4782 \\
05 &   0 & 40 & 0.4 &  93.33 & 0.2826 \\
06 &   0 & 40 & 0.7 & 226.67 & 0.1282 \\
07 &  90 & 30 & 0.1 &  36.67 & 0.5521 \\
08 &  90 & 30 & 0.4 &  70.00 & 0.3263 \\
09 &  90 & 30 & 0.7 & 170.00 & 0.1480 \\
10 &  90 & 40 & 0.1 &  48.89 & 0.4782 \\
11 &  90 & 40 & 0.4 &  93.33 & 0.2826 \\
12 &  90 & 40 & 0.7 & 226.67 & 0.1282 \\
13 &   0 & 15 & 0.1 &  18.33 & 0.7808 \\
14 &   0 & 15 & 0.4 &  35.00 & 0.4614 \\
15 &   0 & 15 & 0.7 &  85.00 & 0.2094 \\
16 &   0 & 20 & 0.1 &  24.44 & 0.6762 \\
17 &   0 & 20 & 0.4 &  46.67 & 0.3996 \\
18 &   0 & 20 & 0.7 & 113.33 & 0.1813 \\
19 &  90 & 15 & 0.1 &  18.33 & 0.7808 \\
20 &  90 & 15 & 0.4 &  35.00 & 0.4614 \\
21 &  90 & 15 & 0.7 &  85.00 & 0.2094 \\
22 &  90 & 20 & 0.1 &  24.44 & 0.6762 \\
23 &  90 & 20 & 0.4 &  46.67 & 0.3996 \\
24 &  90 & 20 & 0.7 & 113.33 & 0.1813 \\
25 &   0 & 23 & 0.1 &  28.11 & 0.6306 \\
26 &   0 & 23 & 0.4 &  53.67 & 0.3726 \\
27 &   0 & 10 & 0.7 &  56.67 & 0.2564 \\
28 &   0 & 23 & 0.1 &  28.11 & 0.6306 \\
29 &   0 & 23 & 0.4 &  53.67 & 0.3726 \\
30 &   0 & 10 & 0.7 &  56.67 & 0.2564 \\
31 &  90 & 23 & 0.1 &  28.11 & 0.6306 \\
32 &  90 & 23 & 0.4 &  53.67 & 0.3726 \\
33 &  90 & 10 & 0.7 &  56.67 & 0.2564 \\
34 &  90 & 23 & 0.1 &  28.11 & 0.6306 \\
35 &  90 & 23 & 0.4 &  53.67 & 0.3726 \\
36 &  90 & 10 & 0.7 &  56.67 & 0.2564 \\
\end{tabular}

\medskip
$G_1$ is the primary galaxy, $G_2=G_1$ the secondary galaxy,
$\Theta$ the angle between the two planes of the discs in units of degree,
$R_p$ the pericentric distance in units of length, $M_1$ the primary galaxy 
mass in units of mass, 
$e$ the eccentricity, $R_a$ the apocentric distance in units of length,
$V_a$ the velocity at the apocentric distance in units of velocity, 
$M_1$ the primary galaxy mass and $M_2=M_1=0.621$ the secondary mass 
galaxy in units of mass.
\end{minipage}
\end{table*}

\begin{table*}
\begin{minipage}{140 mm}
\caption{Characteristics of the Final Stage of the Orbits and Merged Discs}
\label{table3}
\begin{tabular}{@{}cccccccccc}
EXP & Disc Interaction & Number of Orbits & $T_M$ & $R_{d(12)}$ & $Z_{d(12)}$ & $R_{f}$\\
01 &  Open   & 1.5 &      &                  &                  &\\
02 &  Open   & 1.0 &      &                  &                  &\\
03 &  Open   &     &      &                  &                  &\\
04 &  Open   & 1.0 &      &                  &                  &\\
05 &  Open   & 0.5 &      &                  &                  &\\
06 &  Open   &     &      &                  &                  &\\
07 &  Open   & 1.5 &      &                  &                  &\\
08 &  Open   & 1.0 &      &                  &                  &\\
09 &  Open   &     &      &                  &                  &\\
10 &  Open   & 1.0 &      &                  &                  &\\
11 &  Open   & 0.5 &      &                  &                  &\\
12 &  Open   &     &      &                  &                  &\\
13 & Merge   & 1.0 & 0.21 & $0.867\pm 0.041$ & $0.116\pm 0.006$ & 10\\
14 & Merge   & 1.5 & 0.42 & $0.946\pm 0.051$ & $0.142\pm 0.010$ & 10\\
15 & Graze   & 1.0 & 1.60*&                  &                  &\\
16 & Merge   & 1.5 & 0.42 & $0.771\pm 0.042$ & $0.128\pm 0.009$ & 10\\
17 & Merge   & 2.5 & 1.00 & $0.682\pm 0.025$ & $0.199\pm 0.018$ & 10\\
18 &  Open   & 0.5 &      &                  &                  &\\
19 & Merge   & 1.0 & 0.25 &                  &                  &\\
20 & Merge   & 1.5 & 0.43 &                  &                  &\\
21 & Graze   & 1.0 & 1.61*&                  &                  &\\
22 & Merge   & 1.5 & 0.42 &                  &                  &\\
23 & Merge   & 2.5 & 1.00 &                  &                  &\\
24 &  Open   & 0.5 &      &                  &                  &\\
25 & Merge   & 1.5 & 0.63 & $0.791\pm 0.038$ & $0.116\pm 0.006$ & 10\\
26 &  Open   & 1.5 &      &                  &                  &\\
27 & Merge   & 1.5 & 0.50 & $0.871\pm 0.044$ & $0.465\pm 0.079$ & 10\\
28 & Merge   & 1.5 & 0.60 & $0.791\pm 0.038$ & $0.116\pm 0.006$ & 10\\
29 &  Open   & 1.5 &      &                  &                  &\\
30 & Merge   & 1.5 & 0.50 & $0.809\pm 0.041$ & $0.175\pm 0.007$ & 10\\
31 & Merge   & 1.5 & 0.60 &                  &                  &\\
32 &  Open   & 1.5 &      &                  &                  &\\
33 & Merge   & 1.5 & 0.54 &                  &                  &\\
34 & Merge   & 1.5 & 0.60 &                  &                  &\\
35 &  Open   & 1.5 &      &                  &                  &\\
36 & Merge   & 1.5 & 0.54 &                  &                  &\\
\end{tabular}

\medskip
$Open$ means that the two discs do not touch each other during the time of the
experiment ($t_H$). $Graze$ means that the two discs touch each other for a 
while and after they separate.  $Merge$ means that the two discs fuse to each 
other.
{
Number of orbits is the total angular excursion of the companion up to the 
merger, relative to its starting point.
}
$T_M$ is the time of merging in units of $t_H$ when the two discs fuse to each
other (the symbols * in the times of merging of the simulations EXP15 and EXP21
denote that these times are estimations, using EXP17).
$R_{d(12)}$, $Z_{d(12)}$ and $R_{f}$ are the fitted scale radius, height and 
cutoff fitting radius of the unique merged coplanar disc in units of length, 
respectively, using Equation (\ref{dens}).
\end{minipage}
\end{table*}

\begin{figure}
\centering
\includegraphics[width=8cm]{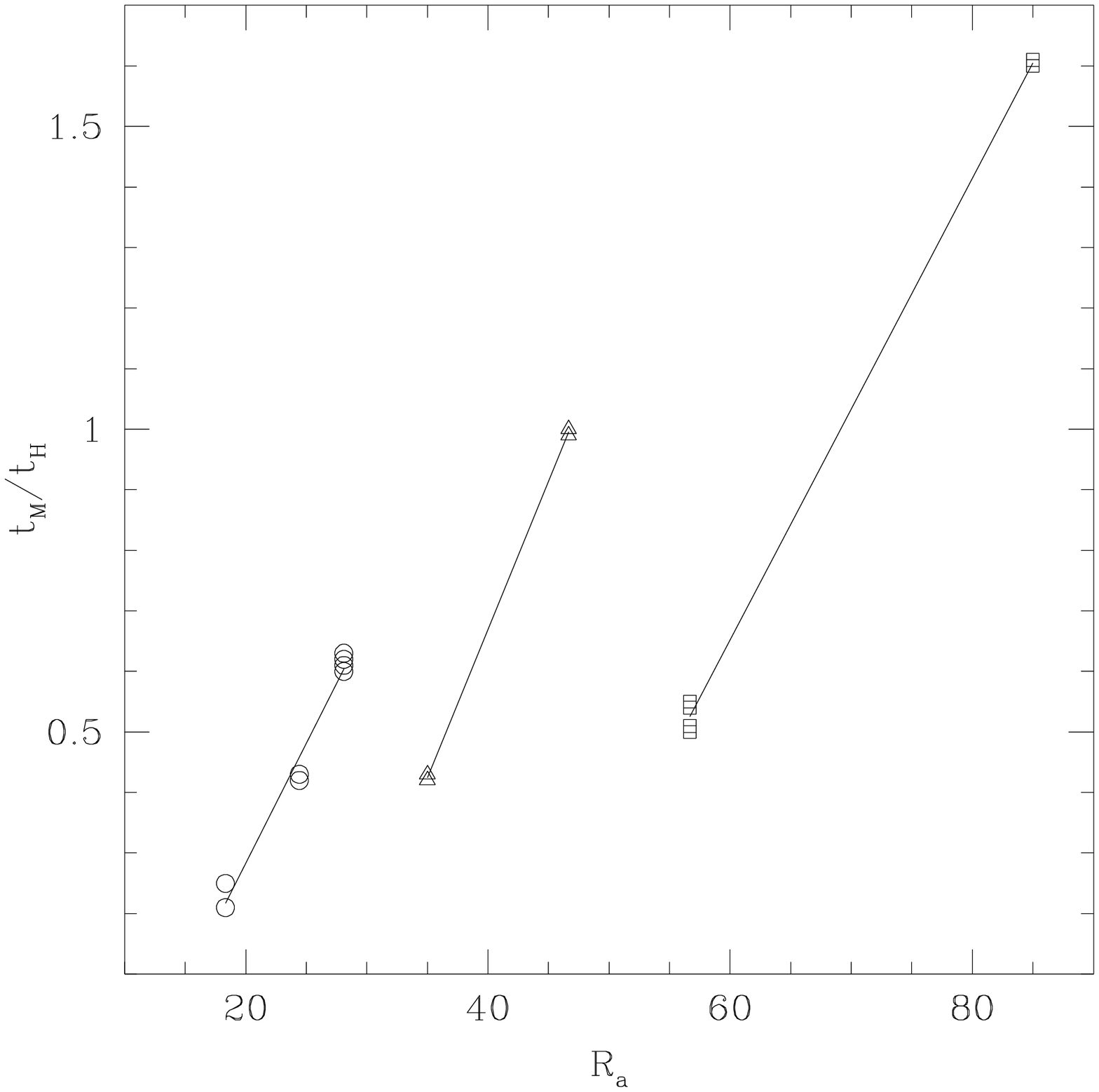}
\caption
{The time of merging with the fitted straight lines, for each
eccentricity, where $R_a$ is the apocentric distance. The open circles
represent the simulations with $e=0.1$. The open triangles represent the
simulations with $e=0.4$. The open squares denote the experiments with $e=0.7$.
The best fit parameters are:
$[t_M/t_H] = (0.039 \pm 0.002) R_a + (-0.508 \pm 0.056)$ (for $e=0.1$), 
$[t_M/t_H] = (0.049 \pm 0.005) R_a + (-1.285 \pm 0.020)$ (for $e=0.4$) and  
$[t_M/t_H] = (0.038 \pm 0.006) R_a + (-1.635 \pm 0.038)$ (for $e=0.7$) 
(the far two points were obtained extrapolating the time 
evolution of the distance between the two discs, using the simulation EXP17).}
\label{merge}
\end{figure}

In Figure \ref{merge} we show the dependence of the time of merging ($T_M$)
(see Table \ref{table3})
with the eccentricity ($e$) and the initial apocentric distance ($R_a$).
The time of merging is defined as the time when the centers of mass of the
two discs (primary and secondary galaxies) overlap each other \cite{Chan2001}.
We notice that the merging time increases linearly with the initial
apocentric distance.  The $T_M$ for different eccentricities is 
obtained extrapolating the linear approximation for each eccentricity
and determined $T_M$ for a fixed value of $R_a$. We have obtained that
the time of merging decreases with eccentricity.

\begin{figure}
\centering
\includegraphics[width=8cm]{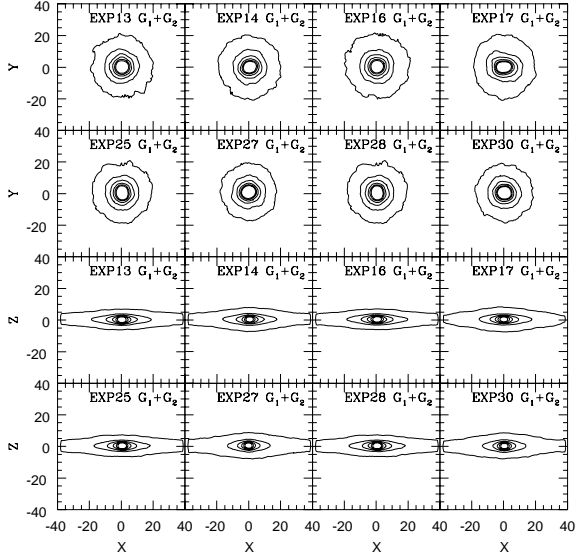}
\caption
{The contour snapshot of the merger of the primary and
secondary galaxies { (flat disk merged remnants)} 
together in the planes XY and XZ,
at the Hubble time of the simulation ($t=t_H$). Simulations EXP13, 14, 
16, 17, 25, 27, 28 and 30.}
\label{snapshot_100a}
\end{figure}

{ There are two different kinds of merged remnants} in our simulations.
One of them (coplanar discs) is a disc galaxy with scale radius and height 
very similar
to the initial disc galaxy (see Table \ref{table3}), but with a tidal radius 
that is at least five times greater
than of the initial galaxy (see Figure \ref{snapshot_100a}).  
The other one (polar discs) resembles 
a lenticular galaxy, but again with a tidal radius that is greater
than the initial galaxy radius (see Figure \ref{snapshot_100a}).

\begin{figure}
\centering
\includegraphics[width=8cm]{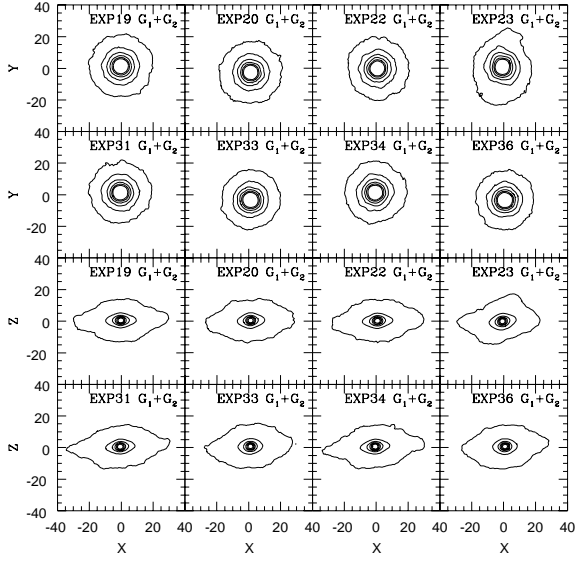}
\caption
{The contour snapshot of the merger of the primary and
secondary galaxies { (oblate disk merged remnants)} 
together in the planes XY and XZ,
at the Hubble time of the simulation ($t=t_H$). Simulations EXP19, 20, 
22, 23, 31, 33, 34 and 36.}
\label{snapshot_100b}
\end{figure}

In the Figures \ref{snapshot_100a} and \ref{snapshot_100b} we present
the contour snapshots of the result of the merger of the primary and
secondary galaxies together in the planes XY and XZ,
at the Hubble time of the simulation ($t=t_H$).  We show the simulations:
EXP13, 14, 16, 17, 19, 20, 22, 23, 25, 27,
28, 30, 31, 33, 34 and 36.
We notice that in the simulations with coplanar disc orbits (EXP13, 14, 16,
17, 25, 27, 28 and 30) the resulting fused galaxies are still disc galaxies.
Their fitted scale radii ($R_{d(12)}$) and heights ($Z_{d(12)}$) using the 
Equation 
(\ref{dens}) are presented in Table \ref{table3}.  We can see that the merged
disc galaxies are thicker and bigger than the initial ones.

However, for the simulations with polar disc orbits (EXP19, 20, 22, 23,
31, 33, 34 and 36) the resulting fused galaxies are not disc galaxies anymore.
In both planes, XY and XZ, the galaxies resemble to lenticular galaxies.
The outer contour level of the merged galaxy in EXP23 is clearly deformed,
differently of others merged polar discs, maybe because of the number
of orbits (see Table \ref{table3}).  This is the unique simulation among
all our experiments with the maximum number of orbits (2.5), i.e., with
maximum interval of time with tidal interaction between the two disc galaxies.

\begin{figure}
\centering
\includegraphics[width=8cm]{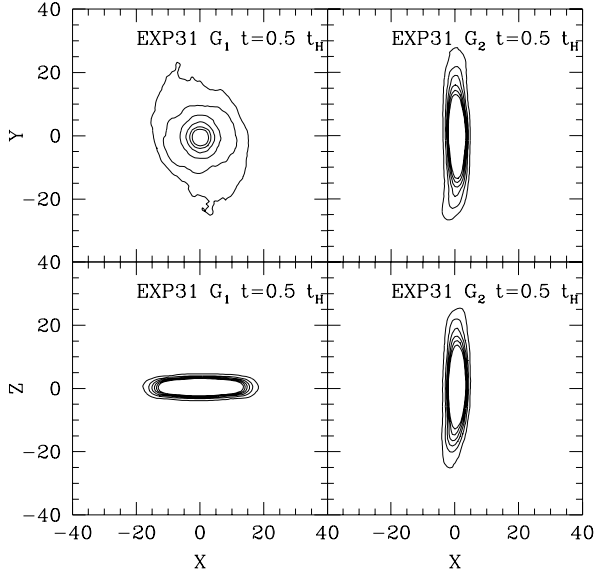}
\caption
{The contour of the snapshot of the merger of the primary and
secondary galaxies plotted separately in the plane XY and XZ,
at 50\% of the Hubble time ($t=0.5 t_H$). Simulation EXP31.}
\label{snapshot_050c}
\end{figure}

\begin{figure}
\centering
\includegraphics[width=8cm]{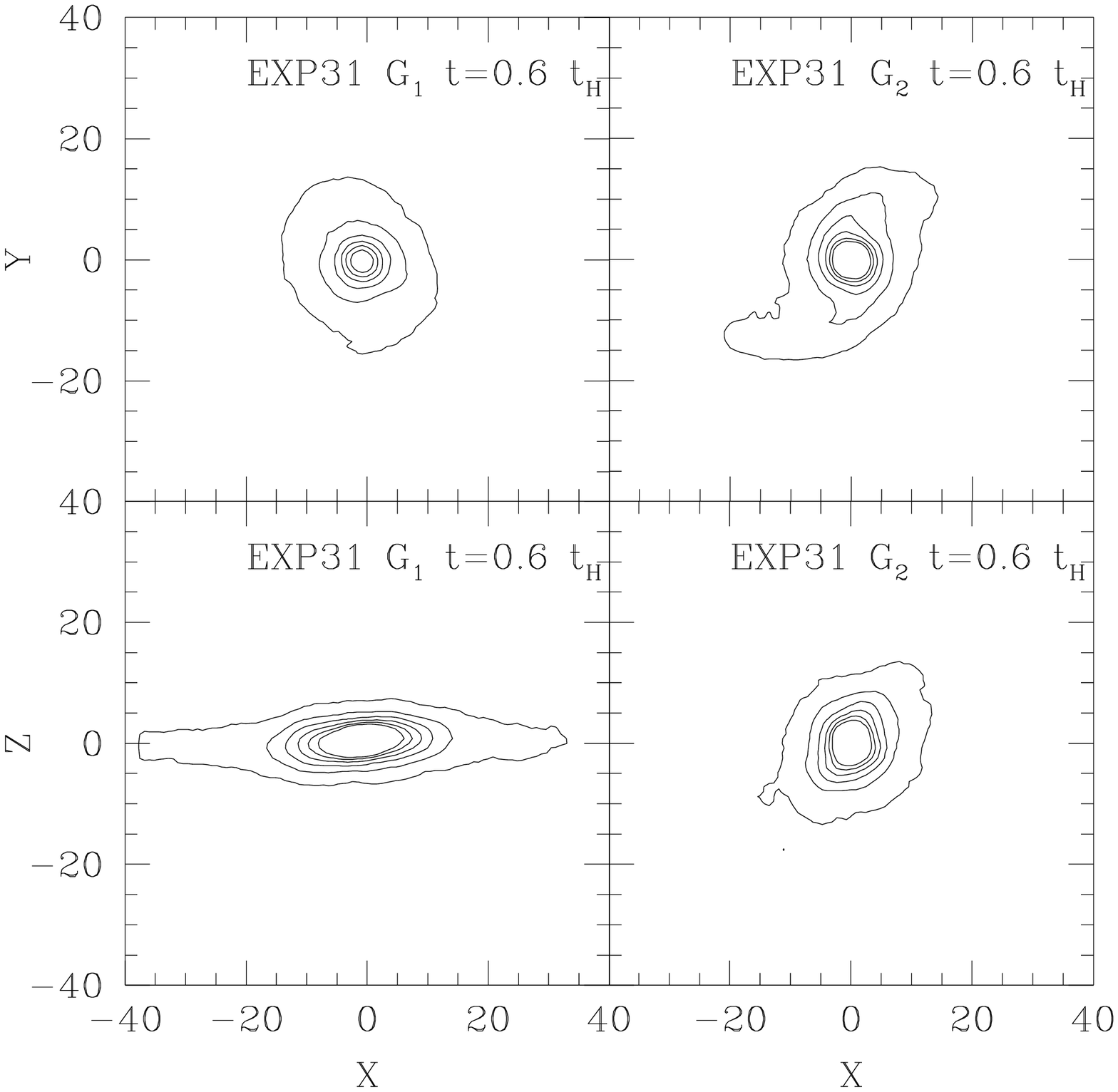}
\caption
{The contour of the snapshot of the merger of the primary and
secondary galaxies plotted separately in the plane XY and XZ,
at 60\% of the Hubble time 
($t=0.6 t_H$). Overlapping the contours
of $G_1$ and $G_2$, at $t=t_H$, we get the contours of Figure 
\ref{snapshot_100b} for the simulation EXP31.}
\label{snapshot_100d}
\end{figure}

In Figures \ref{snapshot_050c} and \ref{snapshot_100d} we can show what
happened to these galaxies using the simulation EXP31 at $t=0.5t_H$ and
at $t=t_H$.  These figures show the discs of the primary galaxy $G_1$ and 
secondary $G_2$.  We can see that the polar characteristic of the $G_2$ is still
there at $t=0.5t_H$ but this is lost at $t=t_H$.  At this time the polar disc
is completely disrupted and its debris form a stellar halo.
Overlapping the contours
of $G_1$ and $G_2$, we get Figure \ref{snapshot_100b} for the
simulation EXP31.

\section{Power Spectrum Analysis}

{
The power spectrum analysis provides a useful and objective tool for the study 
of the induced waves.
This analysis uses the amplitude and the phase of the Fourier components
of the surface density of the stars, allow us to evidence the presence of the 
spiral modes in our simulations.  This analysis give us the pattern speed of 
all present spiral perturbations and their relative positions.
}

The method of power spectrum, known historically as periodogram, is used to 
search for periodicities in sparse, noisy unevenly spaced data
(Junqueira \& Combes 1996).

If we take a N-point sample of the function $c(t)$ at equal intervals of time 
$t$ and compute its discrete Fourier transform (Press at al. 1992)
we get the power spectrum $P(\Omega)$ of $c(t)$.

We have used the grid expansion method in order to analyze the density 
distribution ($128 \times 128 \times 128$ pixels), for obtaining the power
spectrum.

{ 
Firstly, using a radial binning of $0.0125$, we obtain the amplitude and 
the phase of the Fourier components. Secondly, using the snapshots of the slab 
of the disc density 
in a interval of time of $0.01 t_H$, we calculate the superposition of all the 
Fourier amplitude and phase for each radial bin and for each interval of time.
Finally, we obtain the power spectrum as a plot of number density for each 
radial bin.
}
The power spectrum analysis was done by studying the primary galaxy just before
the merging time. The orientation of 
the disc was not followed dynamically because it has not deviated from the 
initial orientation angle, as we can see in Figure \ref{snapshot_100d}.

Since we have a 3D particle disc, we limited the number of the particles 
within the 
planes $Z=-Z_{max}$ and $Z=Z_{max}$ in order to simplify the application of the 
grid expansion method \cite{Chan2003}.  We 
have considered
this thin slab between these two planes as the plane $Z=0$ for the 
grid expansion. { Henceforth}, this thin slab will be denoted as $Z=0$ in the
equations.  The chosen quantity $Z_{max}=0.1$ 
is the value of the scale height of the disc ($Z_d$). 
There are approximately 40\% of the total disc particles ($N_d$) within these 
two planes.  
In all the analysis hereinafter it is assumed a maximum radius of 5 length
units since we have 95\% of the mass of the disc within this radius.

The basic assumption of the density wave theory is that spiral arms are not
always composed of the same stars but instead they are the
manifestation of the excess matter associated with the crest of a rotating
wave pattern.  Two further assumptions were introduced from the onset,
the linearity and quasi-stationarity of the wave.  These assumptions
allow us to write any perturbation of the axisymmetric background as 
superposition waves given by    

\begin{equation}
\rho_d(R,\phi,Z=0,t)=\sum_m \rho_m(R) e^{i[\Omega(m) t - m\phi]},
\label{rho}
\end{equation}
where ${ \rho_d}$ is the density.  The
summation index indicates the symmetry of the component:  $m=0$ corresponds to
the axisymmetric background; $m=1$ corresponds to the lopsided perturbation and 
$m=2$ corresponds to the symmetric two arms perturbation (spiral, bar).  
$\Omega(m)$ is the pattern speed of the component $m$.

We can rewrite Equation (\ref{rho}) in the usual wave 
notation
\begin{equation}
\rho_d(R,\phi,Z=0,t)=\sum_m p_m(R) e^{i[\Psi_m(R) + \Omega(m) t - m\phi]},
\label{rho1}
\end{equation}
where $p_m(R)$ is the amplitude of the wave and $\Psi_m(R)$
is the phase angle of the mode $m$.

Now the density mode $m$ can be obtained from

\begin{equation}
\rho_d^m(R,\phi,Z=0,t)= p_m(R) \cos[ \Psi_m(R) + \Omega(m) t - m\phi ].
\label{rhodmr}
\end{equation}

\begin{figure}
\centering
\includegraphics[width=8cm]{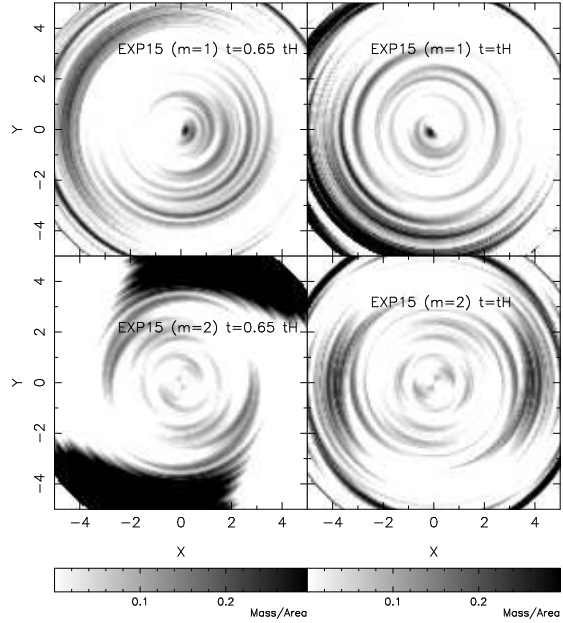}
\caption
{The wave modes $m=1$ and $m=2$  
for the simulation EXP15,
at two different instants of time $0.65 t_H$ and $t_H$. 
The density levels for these plots are the same used in $m=1$ ($t=0.65 t_H$).}
\label{pconstm1m2exp15}
\end{figure}

\begin{figure}
\centering
\includegraphics[width=8cm]{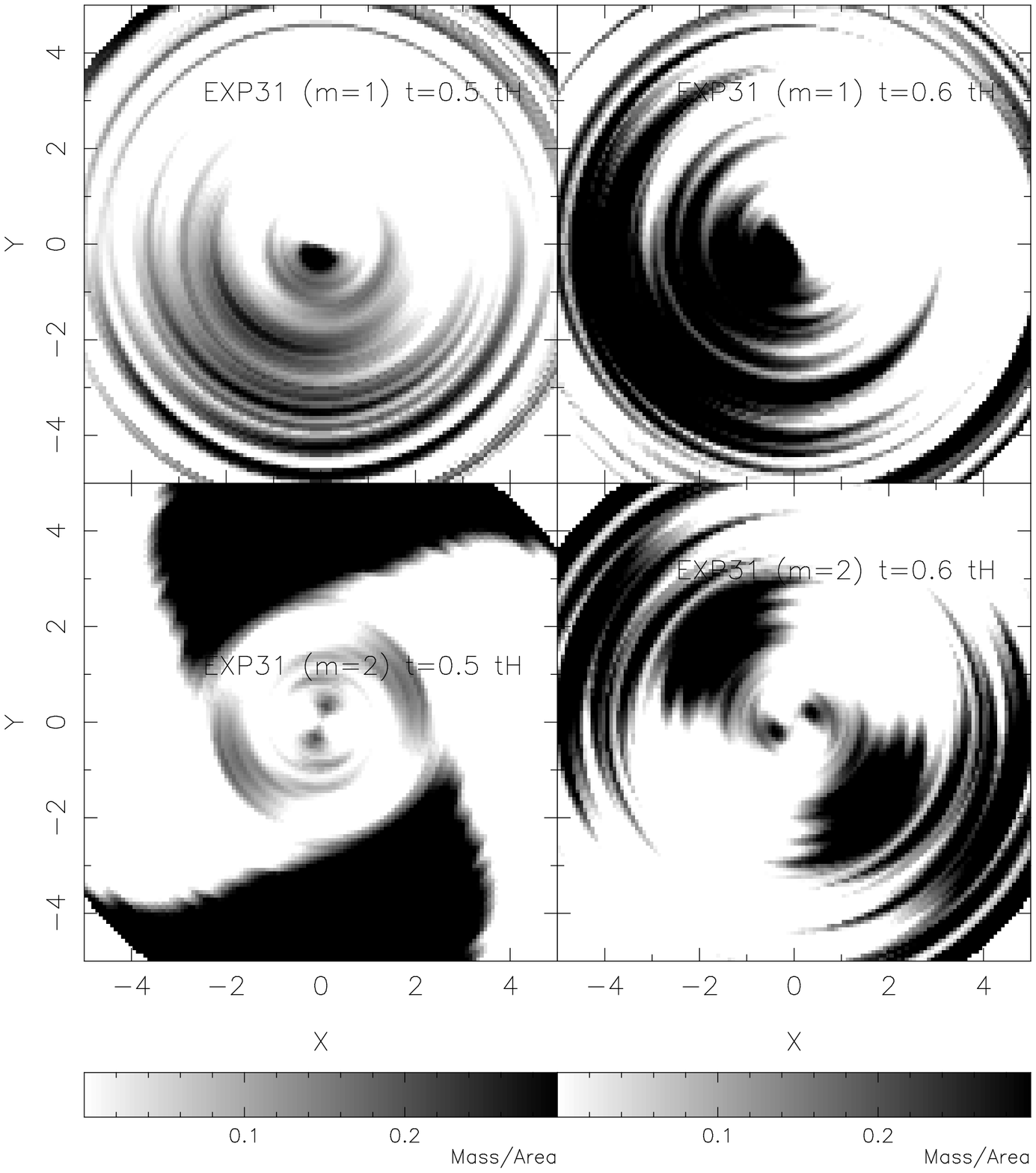}
\caption
{The wave modes $m=1$ and $m=2$  
for the simulation EXP31,
at two different instants of time $0.5 t_H$ and 
$0.6 t_H$. 
The density levels for these plots are the same used in EXP15 
($m=1$, $t=0.65 t_H$).}
\label{pconstm1m2exp31}
\end{figure}

{
Let us interpret the $m=1$ and $m=2$ plots in Figures \ref{pconstm1m2exp15} 
and \ref{pconstm1m2exp31}.
They do not look like a clear one-armed spirals, or a clear asymmetry in the 
two-arms, like in the Figure 13 of the work of Junqueira \& Combes 
\cite{Junqueira1996}, because in this present work
they represent the Fourier analysis of a transient wave.
Junqueira \& Combes analyzed only the gaseous disc, but the analysis is similar
to a stellar disc.
}

In Figure \ref{pconstm1m2exp15} we show 
the transient wave modes $m=1$ and $m=2$ for the simulation EXP15,
at two different instants of time $0.65 t_H$ and $t_H$.  
We notice that transient $m=1$
wave modes are mostly present in outer part of the discs.
{ We note also that transient spiral arms ($m = 2$) are
formed in the outer regions of the discs, and bars are present in the 
inner regions}.
Since they are transient $m=2$ wave modes
the power spectrum for EXP15 (see Figure \ref{powm2}) shows an undefined
angular velocity for this mode.

\begin{figure}
\centering
\includegraphics[width=8cm]{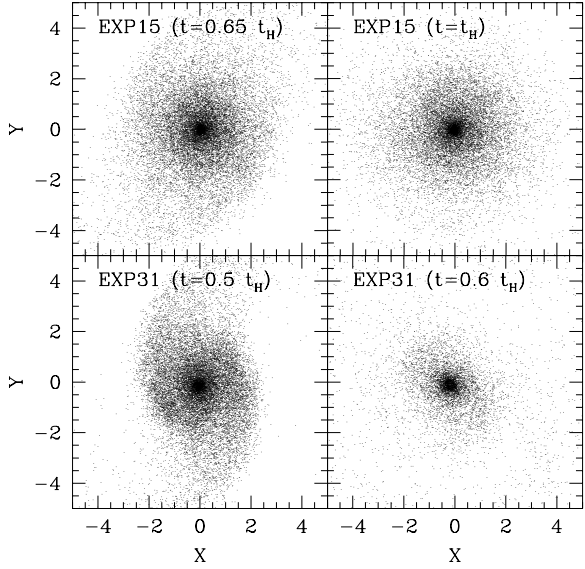}
\caption
{{ The snapshots of the slabs $Z=0$ for the simulation EXP15,
at two different instants of time $0.65 t_H$ and $t_H$. 
Also, the snapshots of the slabs $Z=0$ for the simulation EXP31,
at two different instants of time $0.5 t_H$ and $0.6 t_H$.}} 
\label{pxydiscexp15_31}
\end{figure}

\begin{figure}
\centering
\includegraphics[width=10cm]{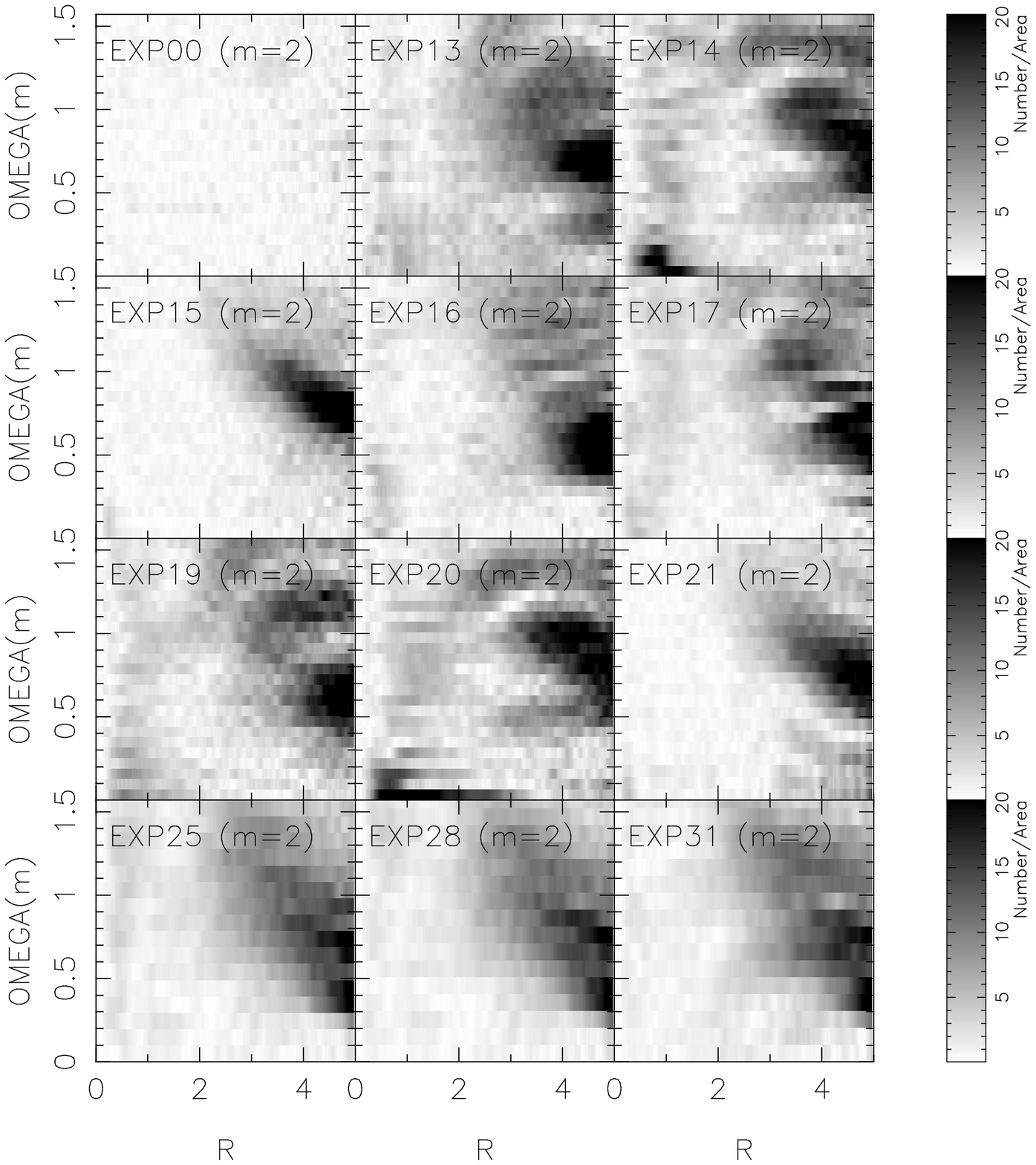}
\caption
{The power spectrum for the mode $m=2$ for the primary galaxy ($G_1$)
and for the simulations EXP00, 13, 
14, 15, 16, 17, 19, 20, 21, 25, 28 and 31 
until the time of merging.
The density levels are three times greater than that of the 
experiment EXP00.}
\label{powm2}
\end{figure}

In Figure \ref{pconstm1m2exp31} we show 
the transient wave modes $m=1$ and $m=2$ for the simulation EXP31,
at two different instants of time $0.5 t_H$ and $t_H$.  
We notice that the transient $m=1$
wave mode at $t=0.5 t_H$ is mostly present in outer part of the discs,
except at $t=t_H$.
There is a big transient spiral arm ($m=2$) at $t=0.5 t_H$
in the outer region of the disc and a { prominent} bar  
in the inner region of the disc at 
$t=0.6 t_H$.  
As in the EXP15, here we have transient $m=2$ wave modes the power spectrum 
for EXP31 
(see Figure \ref{powm2}).
{
If we compare the Figures \ref{pconstm1m2exp31} and \ref{pxydiscexp15_31}
we can see the Fourier decompositions are very similar with the snaphots
of the discs.
}

In Figures \ref{powm2} we can see the power 
spectra for the $m=2$ wave mode for the simulations EXP00, 13, 14, 15, 16,
17, 19, 20, 21, 25, 28 and 31.  We have shown only the $m=2$ wave mode because 
we have not detected any $m=1$ wave mode in any simulation.
Others experiments that presented $m=2$ wave mode and that are
not shown in this figure are: EXP22, 23, 27 and 33.  These simulations had
similar behaviors as shown in Figure \ref{powm2}. Others simulations
have not shown any sign of $m=2$ wave mode mostly because the primary and
secondary halo did not touch each other during their evolution time
(open disc interaction: see Table \ref{table3}).

In Figure \ref{powm2}, the first plot shows the power spectrum for the mode 
$m=2$, for the simulation EXP00, without the secondary galaxy.  
This was done in order to analyze the 
existence of self-excited gravitational instabilities  
$m=2$ wave mode in the disc. As we can see, there are not any wave modes.

Note that in Figure \ref{powm2} 
the fuzzy small perturbations in the outer radii of the discs
(note also that the density levels are three times greater than that used 
in the experiment EXP00).
Most of the experiments in this figure have shown merged discs
(see Table \ref{table3}), except the simulations EXP15 and 21 (grazing discs).
There we can also see partial $m=2$ wave modes in the outer radii of the disc 
with 
high clumpy density regions that do not stretch to the inner part of the discs.
Thus, we cannot classify them as being $m=2$ stable wave modes because these
characteristics of these power spectra.

{
There are not discernable and significant substructures in Figure \ref{powm2}.
If we had a real stable wave we should have a plot like the Figure 16 of the
paper of Junqueira \& Combes \cite{Junqueira1996}, a straight line parallel to
the radial axis giving the
pattern speed of the mode.  Instead, since we have a transient wave, we get a
fuzzy plot, more concentrated in the outer radial regions.
}

\begin{figure}
\centering
\includegraphics[width=9cm]{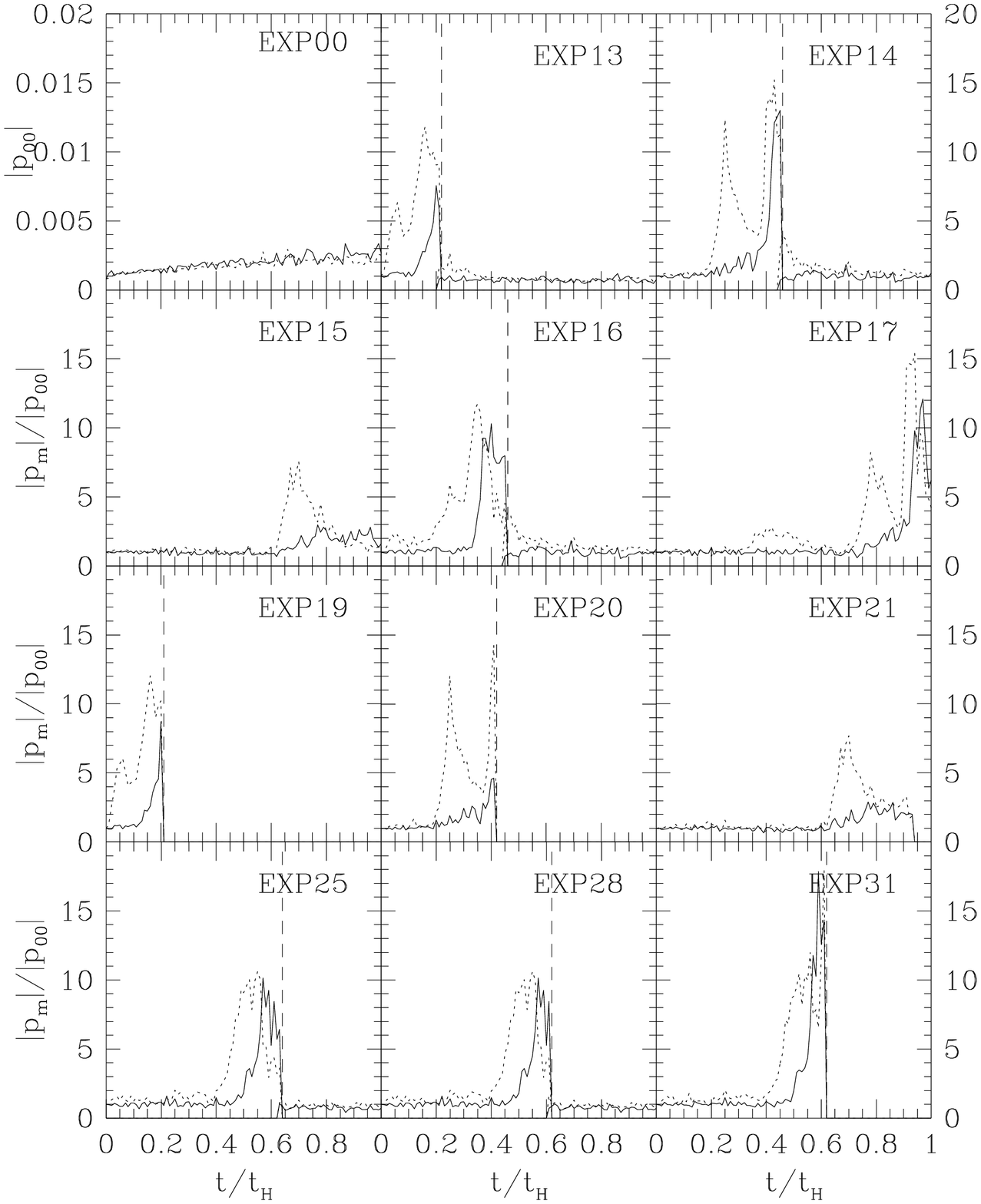}
\caption
{The global integrated { ratio} 
amplitudes of the modes $m=1$ (solid 
lines) and $m=2$ (dotted lines) for the 
primary galaxy ($G_1$) for the simulations 15 and 17 during a Hubble 
time.  { The $|p_{00}|$ denotes the global amplitude of the control 
simulation EXP00}.
For the simulations EXP13, 16, 25 and 28 we show the global { ratio} 
amplitudes
for $G_1$ until the merger time, denoted by the dashed lines.  After the merge
we show the global { ratio} amplitude of the compound galaxy ($G_1+G_2$).
For the experiments EXP19, 20 and 31 (polar disc), we plot the evolution 
of $G_1$ until the merger of $G_1$ and $G_2$, represented again by the dashed
lines.} 
\label{amplitudes}
\end{figure}

We have integrated the wave amplitudes radially to get the global
Fourier amplitudes $|p_m|$ 
\cite{Harsono2011}, where m is the Fourier mode:
\begin{equation}
|p_m|(t) = \frac{1}{N_{disc}}\sum_{R=0}^{R=5} |p_m(R) e^{i[\Psi_m(R) + \Omega(m) t - m\phi]}|,
\end{equation}
where $p_m(R)$ is giving by equation \ref{rho1} and $N_{disc}$ is total number
of particles in the disc within $R=5$.
The global amplitude analysis was done in two distinct ways: a) studying the
primary galaxy until the merging time, b) analyzing the compound galaxies after
the merge, when they occur.  The orientation of the disc was not followed
dynamically because of the deformation of the disc in some simulations.
In Figure \ref{amplitudes} we show the temporal evolution of the { relative}
global
integrated amplitudes for the simulations EXP13, 14, 15, 16, 17, 19, 20, 21,
25, 28 and 31.

As we can see in Figure \ref{amplitudes}, there is not a possible kick due to 
non cosmological initial conditions.  The reason is simply because  all the
simulations have begun at the apocentric distance, where the tidal
interaction between the two binary galaxies is weaker.

{
Furthermore, all the waves are mostly driven shortly before merger, because the 
separations have gotten small, even in the cases with no merger.

Let us analyze the special case EXP17 in Figure \ref{amplitudes}. 
The simulation EXP17 presents three maxima for $m=2$ wave mode. The first 
maximum is due to the formation of a bar, when the two discs are still far 
apart.  The second and third ones are due to the formation of a two-arms spiral,
when the discs are already touching and deforming each other.
}

{ Lopsided features are preferentially observed
in the distribution of gas in late-type spiral galaxies.
In several cases these features can be identified as one-armed
spirals ($m=1$ mode). More frequently, nuclei of galaxies are observed
displaced with respect to the gravity center, as in M33 and M101.
The nucleus of M31 reveals such an off-centering which
has been interpreted in terms of an $m=1$ perturbation.
Miller and Smith \cite{Miller1992} have studied through N-body simulations of
disk galaxies, a peculiar oscillatory motion of the nucleus with respect
to the rest of the axisymmetric galaxy. They interpret the phenomenon
as a $m=1$ instability, a density wave in orbital motion around the center
of mass of the galaxy. Moreover, Junqueira and Combes \cite{Junqueira1996} 
have shown
that stars and gas are off-centered with respect to the center of mass of the 
system. In Figure \ref{amplitudes} we can note that the $m=1$ and $m=2$ modes 
are excited at different times.  This is due to the fact that the mode
excitation comes from the outer region to the inner region of the disc.
Thus, there is a time lag in order to this excitation reaches the inner disc 
region.
Since the $m=1$ mode needs an off-centered disc mode with respect to the center 
of mass and since the $m=2$ mode is mostly excited from the outer disc region, 
this produces a time delay among the maxima of these two modes.} 

\begin{figure}
\centering
\includegraphics[width=9cm]{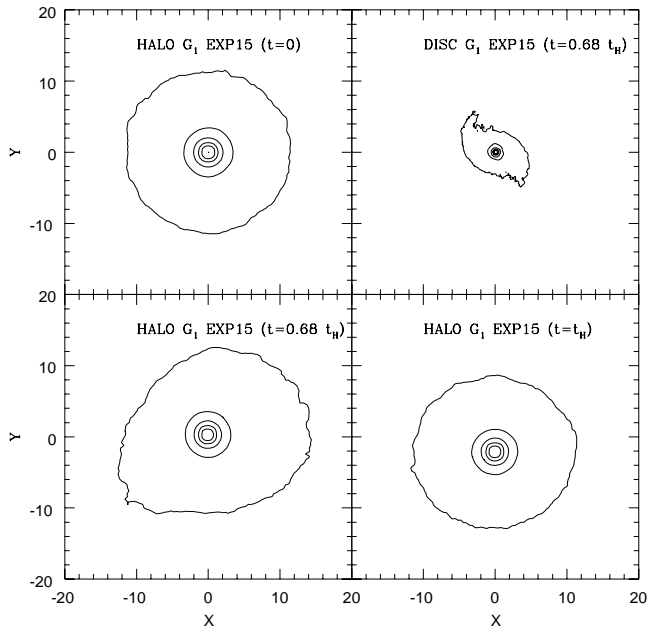}
\caption
{{ The $G_1$ halo contours of the EXP15 in the plane XY at three different 
times: $t=0$, $t=0.68 t_H$  at the time of the maximum amplitude of 
the disc $m=2$ component (see Figure \ref{amplitudes}), and $t=t_H$. 
The $G_1$ disc contour of the EXP15 in the plane XY at the time $t=0.68 t_H$.
The two outer halo contour density levels correspond approximately to 40\% of 
the total halo 
mass at the radius $R\approx 4$ and 90\% of the total halo mass at the radius 
$R\approx 11$.
The halo and the disc of the secondary galaxy $G_2$ can be obtained just 
reflecting the respective
contour image relative to the $Y$ axis at  $X=0$, since the two galaxies have
the same mass distribution.}}
\label{contorxyhalo_exp15}
\end{figure}

\begin{figure}
\centering
\includegraphics[width=9cm]{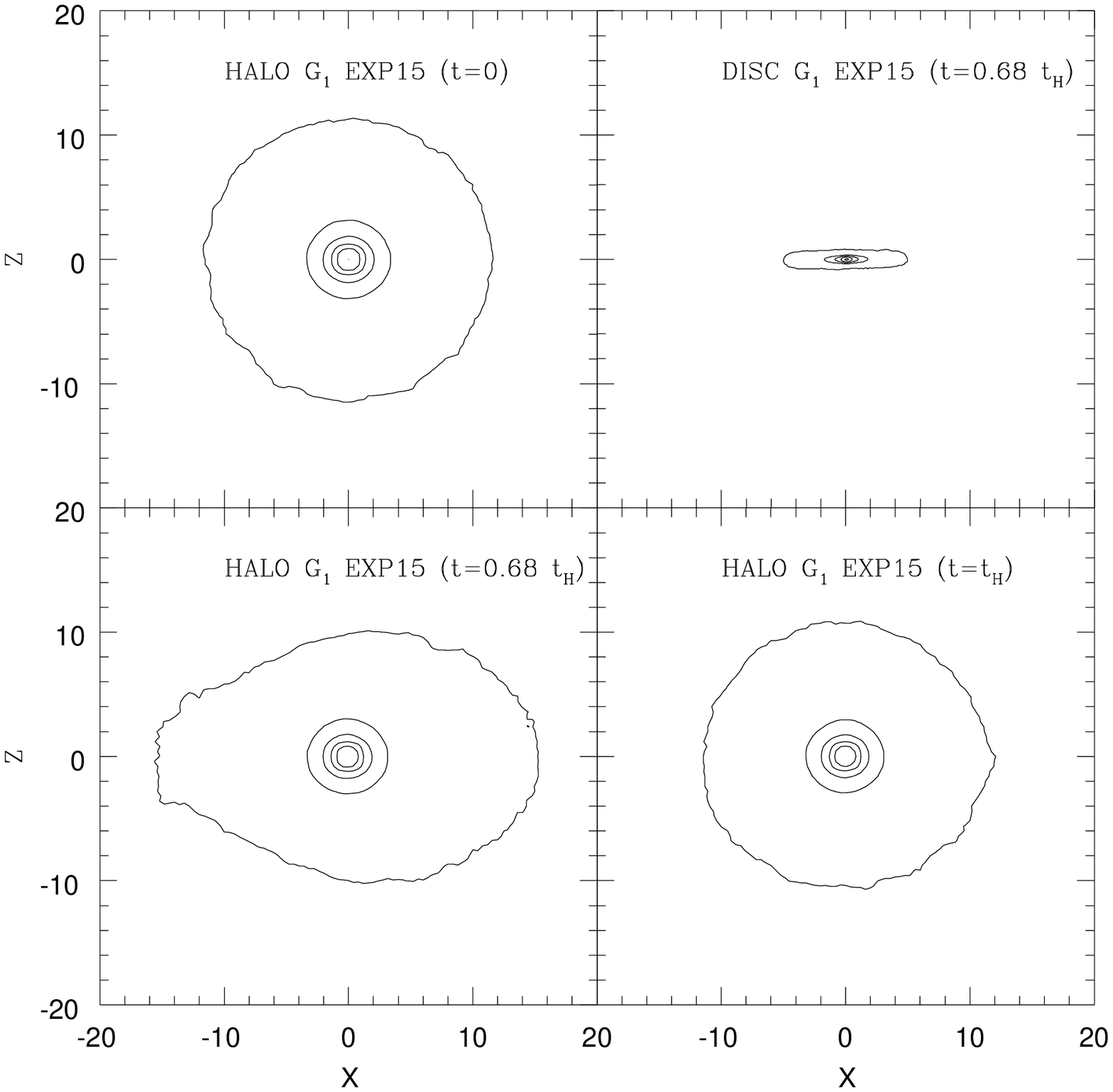}
\caption
{{ The $G_1$ halo contours of the EXP15 in the plane XZ at three different 
times: $t=0$, $t=0.68 t_H$  at the time of the maximum amplitude of 
the disc $m=2$ component (see Figure \ref{amplitudes}), and $t=t_H$. 
The $G_1$ disc contour of the EXP15 in the plane XZ at the time $t=0.68 t_H$.
The two outer halo contour density levels correspond approximately to 40\% of 
the total halo 
mass at the radius $R\approx 4$ and 90\% of the total halo mass at the radius 
$R\approx 11$.
The halo and the disc of the secondary galaxy $G_2$ can be obtained just 
reflecting the respective
contour image relative to the $Y$ axis at  $X=0$, since the two galaxies have
the same mass distribution.}}
\label{contorxzhalo_exp15}
\end{figure}

{
The Figures \ref{contorxyhalo_exp15} and \ref{contorxzhalo_exp15} show the 
evolution of some halo contours containing, for example, about 40\% and 90\% 
the halo mass 
(EXP15). The early merger paper of Barnes and Hernquist \cite{Barnes1996}, 
which first discussed about the halo mergers, has given attention only to what
happened with the remnant halos at the end of the simulation. As we can see in 
the present Figures, the maximum
halo contour deformation (90\%) coincides with the maximum disc deformation.
After the passage of the secondary galaxy through the primary at the Hubble 
time, the halos settles down and their contours resemble with the initial ones.
}

\section{Discussion}

We have evolved dynamically, using N-body simulations, 
two disc galaxies with halo and
bulge.  The initial disc model is stable against any self-excited $m=1$
or $m=2$ wave modes.  
{
The satellite galaxy has coplanar or polar disc orientation in relation
to the disc of the primary galaxy and their initial orbits are prograde
eccentric ($e=0.1$, $e=0.4$ or $e=0.7$).
}
Both galaxies have similar mass and size of the Milk Way.

Most of the recent papers that studied the tidal interaction between two
galaxies have used a fixed potential for the halo \cite{Oh2008,Dobbs2010,
Struck2011}.  This condition can mislead the results because the { live 
halos}
are very important to transmit angular momentum to the disc of the primary
galaxy.  The halo of the primary galaxy can respond globally to disturbance of
the halo of the secondary galaxy, thus it can affect the disc structure
in an inward effect. These effects can be clearly seen in the analysis of
the power spectra (see Figure \ref{powm2}).

{ We note that this is the first published work, as far as we know,}
that has studied the secular evolution of bound disc binary galaxies.
Nevertheless, we will only compare our results with the global results of
similar papers, since the numerical methods, initial conditions, time of
integration, etc., are different from ours.

We have shown that the merger of two 
coplanar ($\Theta=0$) 
{ pure stellar} disc galaxies can result in a disc galaxy,
instead of an elliptical one, as it is shown in { other papers}
\cite{Bournaud2005,Bois2011}. 
If we have the merger of two polar ($\Theta=90$) disc galaxies 
we can also have formation of
lenticular-like galaxies.  These results are new in the literature, as far as we
have knowledge.  

In fact, none of our simulations resulted in elliptical galaxies.
In a recent work Bois et al. (2011) has studied the formation of early-type 
galaxies through mergers with a sample of high-resolution numerical simulations 
of binary mergers of disc galaxies. 
The initial galaxy model had alive halo, bulge, disc and gas.
The orbits used in the merge
simulations were all parabolic or hyperbolic, corresponding to initially
unbound galaxy pairs, differently of our simulations where the galaxy pairs
were, from the very beginning, bound in eccentric orbits.

{ Furthermore}, we have demonstrated that the time of merging increases 
linearly with
the initial apocentric distance of the galaxies and decreases with the
eccentricity (see Figure \ref{merge}).
In their paper Boylan-Kolchin \& Quataert (2008) have studied the merging time
 of extended dark matter haloes using N-body simulations.
Each of their simulations consists of a host halo and a satellite halo; 
the ratio of the satellite to the host mass, varied from 0.025 to 0.3 and 
initial circularity of the satellite varied from 0.33 to 1, i.e., the initial
eccentricity varied from 0 to 0.67.
They have found that the merging time decreases exponentially with the 
eccentricity.  This result is in partial agreement with our findings since
the $T_M$ decreases with the eccentricity.  { However},
we do not have enough simulations with different eccentricities to
confirm the exponential behavior.

We also have shown that the tidal forces of the discs can
excite the wave mode $m=1$ and the wave mode $m=2$, but they are not
stable, i.e., they are transient wave modes (see Figure \ref{powm2}).
In a previous work \cite{Chan2003} we have shown that tidal interaction
of a secondary point-mass galaxy could excite stable $m=1$ and $m=2$ wave modes
in the density distribution as well as in the velocity distribution.
{ In contrast to our
previous} paper, here we begin the simulations with an
apocentric distance where the halos do not touch each other.
However, after the merging of the discs, when it has happened, such
instabilities have faded away completely and the fused disc has become
thicker and bigger.

{
Many authors \cite{Oh2008,Lotz2010,Dobbs2010,Struck2011,Snaith2012}
have shown that the tidal interaction can trigger
gravitational instabilities, such as spiral arms or lopsidedness.
Our results have confirmed the results of these papers, that it was  possible 
to create spiral arms, bars or
lopsidedness through the tidal force, but they were transient phenomena.  

The simulations with merger remnants, the waves abruptly disappear after 
the merger is
completed (in less than one outer disc rotation period). This point,
illustrated in Figure \ref{amplitudes},
shows that it is almost
the opposite result of Struck et al. \cite{Struck2011}, who have found that weak
flybys induce waves that take a long time to nonlinearly break.
The maximum relative amplitude of these waves is at most about 15 times greater
compared to the control case.
The $m=2$ wave mode is generated mainly by tidal interaction in the outer
region of
the discs. The $m=1$ wave mode depends mostly of an interaction of the
inner part of the discs, producing an off-centering effect of the wave mode
center relative to the center of
mass of the disc. These characteristics produce a time lag among the maximum
formation
of these two wave modes. The disc settles down quickly after the
merger, in less than one outer disc rotation period.
Furthermore, though the two discs may
spend a long time in orbit, waves are only induced in the short time they are
close together. The stellar discs can survive gentle merging,
even with a massive companion and the waves abruptly disappear after the
merger is completed.
}

{
Finally,
galaxy discs are born gas-rich, and the key to S0 formation is how to get 
there from such progenitors. It is theoretically interesting that some form of 
disc can be preserved through some types of major merger. Practically, however, 
it is not likely that too many S0s are made as a result of S0+S0 or early Sa+Sa 
mergers. A related, and more important point is that if stellar discs can 
survive some gas-free, major mergers, then they are also likely to survive 
multiple, minor mergers, which may play a more important role in finishing
the formation of S0s. The idea that minor mergers play such a role in 
ellipticals is very well known nowadays, making it for S0s is much more
enlightening.
}

\bigskip
\noindent {\bf ACKNOWLEDGMENTS}
\bigskip

One of the authors (RC) 
acknowledges the financial support from FAPERJ 
(no. E-26/171.754/2000, E-26/171.533/2002 and E-26/170.951/2006 
for construction of a cluster of 16 INTEL PENTIUM DUAL CORE PCs) and the 
other author (SJ) also 
acknowledges the financial support from FAPERJ (no. E-26/170.176/2003).
The author (RC) also acknowledges the financial support
from Conselho Nacional de Desenvolvimento Cient\'{\i}fico e Tecnol\'ogico - 
Brazil.

We also would like to thank the generous amount of CPU time given by LNCC
(Laborat\'orio Nacional de Computa\c c\~ao Cient\'{\i}fica),
CESUP/UFRGS (Centro Nacional de Supercomputa\c c\~ao da UFRGS),
CENAPAD/UNICAMP (Centro Nacional de Processamento de Alto Desempenho da 
UNICAMP),
NACAD/COPPE-UFRJ (N\'ucleo de Atendimento de Computa\c c\~ao de Alto
Desempenho da COPPE/UFRJ) in Brazil.
Besides, this research has been supported by SINAPAD/Brazil.

The authors would like to thank Dr. Vladimir Garrido Ortega for the useful
discussions at the very beginning of this work.

{ 
We acknowledge Dr. Curt Struck for the careful reading of the manuscript
and giving many suggestions that improved this work. 
}

\end{document}